\documentclass[12pt]{article}
\usepackage{a4wide}
\usepackage{amsfonts,amsbsy}
\def\be{\begin{equation}}
\def\ee{\end{equation}}
\def\bea{\begin{eqnarray}}
\def\eea{\end{eqnarray}}

\def\vec#1{\boldsymbol{#1}}
\begin{document}
\begin{titlepage}
\begin{flushright}
{\small preprint LPSC-05-113}\\
{\small hep-ph/0511209}
\end{flushright}
\vfill
\begin{center}
{\LARGE \bf Perspectives}\\[5pt]
{\LARGE\bf  in hadron spectroscopy%
\footnote{Invited Talk at the XIth International Conference on Elastic and Diffractive Scattering, Blois, France, May 15--20, 2005, to appear in the Proceedings }}\\[2cm]
{\large J.-M.\ Richard\footnote{{\it e-mail}: Jean-Marc.Richard@lpsc.in2p3.fr }}\\
{\small Laboratoire de Physique Subatomique et Cosmologie}\\[-2pt]
{\small Universit\'e Joseph Fourier--IN2P3-CNRS,}\\[-2pt]
{\small 53, avenue des Martyrs, 38026 Grenoble cedex, France}
\end{center}
\vfill
\begin{abstract}
A brief survey is presented of selected recent results in hadron spectroscopy and
 related theoretical studies. This includes the pentaquarks and hadrons
containing one or two charmed quarks or antiquarks.
\end{abstract}
\vfill
\end{titlepage}

\section{Introduction}
In recent months, several new results on hadron physics came from a variety
of facilities. Some experiments are not primarily devoted to
spectroscopy, but thanks to excellent particle identification and
analysis devices, they can perform very well in this field. The
tentative discovery of pentaquark states was perhaps the most known of
the new findings, but the new excited states  of mesons with open or
hidden charm, and the new baryons with charm $C=1$ and $C=2$ are also of  importance.

\section{Some new results}
Several experimental groups have demonstrated at this Conference the physics potential of their detectors and shown some of their results. This includes,  in particular:
\begin{itemize}\itemsep -2pt
\item low-lying excitations of $D_s$ mesons, the so-called $D_{s,J}^*$ mesons,
\item long-awaited missing states of charmonium,
\item new states near 4 GeV$/c^2$ in the charmonium spectrum,
\item confirmation and extension of our knowledge of the spectrum of
singly-charmed baryons,
\item evidence for doubly-charmed baryons, not yet confirmed by other
experiments,
\item controversial evidence for the exotic baryons with strangeness or anticharm.
\end{itemize}

The new findings are not always due to a brute-force increase of statistics, but
to the use of new decay channels or new production mechanisms.
For example, the radial ($\eta{_c}'$) and orbital ($h_c$) excitations in the
spin-singlet sector or charmonium were searched for, and discovered using double-charm production in $\mathrm{e}^+\mathrm{e}^-$ collisions and weak decay of $\mathrm{B}$ mesons, instead of the  transition from the higher charmonium states. This ends the charmonium-singlet saga \cite{Martin:2003rr}, and  opens new perspectives. For instance, the double-charm mechanism  responsible for the $\eta'_{c}$ detection at Belle, as recoiling
against the $J/\psi$, might in future lead to study double-charm
baryons and (exotic) the double-charm mesons.

Among the new hadron states, some of them are good candidates for exotic
structures:  chiral partners of ground-states,
hybrid mesons (quark, antiquark and constituent gluon), four-quark
states, or meson--meson molecules. We shall briefly review each of
these sectors.

\section{Understanding new hadrons}
\subsection{Hybrids}
Rather early in the quark model  of meson spectroscopy, 
speculations have been made on new type of excitations, beyond the conventional orbital and radial excitations of a non-relativistic two-body system in a potential
\cite{Giles:1976hh}. The spectrum and properties of hybrids have been refined, and advertised, from the studies made in the framework of the flux tube model or lattice QCD.
A remarkable property is that the hybrids, denoted $(q\bar{q}g)$, do
not decay easily into two ground-state mesons, as one of them is preferably an
orbital excitation.

Within the adiabatic bag model and  also in lattice QCD, the ordinary quarkonium
potential corresponds to the slow motion of a heavy $Q\overline{Q}$
pair, with the gluon field remaining in its state of the lowest
energy. If the gluon field is excited, another Born--Opppenheimer
potential is generated, with a new sequence of bound states. This is
very similar to the spectroscopy of $\mathrm{H}_2{}^+$ in atomic
physics, with the ground state and a first sequence of excitations in
which the electron is in its lowest orbital, and other sequences, where
the electron lies in an excited level. 

In the paper revealing the $X(3940)$ \cite{Abe:2004zs}, the Belle collaboration
mentions its possible hybrid nature. Maiani et al.\ \cite{Maiani:2004vq}, Close et al.\ \cite{Close:2005iz}, and Kou et al.\ 
\cite{Kou:2005gt}, among many others,  have recently analysed the latest results in the hidden-charm
sector, and arrive to somewhat different conclusions as to which of the new sates is more likely an hybrid, a four-quark state, or  a mere $(c\bar{c})$ excitation.
\subsection{Diquark models}
It is extremely probable that the hadron spectrum can be understood,
at least in first approximation, in terms of a few effective entities,
in the same way as chemistry use atoms and ions as ingredients whose
inner structure rarely matters.
The constituent quark model is an illustration, with 
strongly interacting and massless quarks and gluons making dressed constituent quarks, which in turn build the spectrum of light hadrons. 
A further simplification consists of assuming that somehow two quarks
made a diquark and that baryons are in first approximation made of a
quark and a diquark.

The diquark model has been used, perhaps too speculatively, to predict
the diquark--antidiquark states, with orbital excitation, as a model of
baryonium. When the baryonium disappeared from the tables, the diquark
model was looked at with scepticism. It was, however, resurrected with
new argumentation, by Jaffe and Wylceck \cite{Jaffe:2003sg}, who saw in the diquark
model a possible solution to the problem of scalar mesons and of
pentaquarks. In this approach, the light pentaquark $\theta^+(1540)$ is
essentially an $\bar{s}$ antiquark surrounded by two $(ud)$ diquarks
in a relative $\ell=1$ state of orbital momentum. Karliner and Lipkin
\cite{Karliner:2003dt} further assumed a $(\bar{s}ud)$ type of triquark clustering with
a relatively low mass. Maiani et al.\ \cite{Maiani:2004vq} explain the $X(3872)$ \cite{Choi:2003ue} and $X(3940)$ \cite{Abe:2004zs} as a
$(c\bar{q})-(\bar{c}q)$ structures, and that of $Y(4260)$ \cite{Aubert:2005rm} as $(c\bar{s})-(\bar{c}s)$.

It should be understood that these sub-structures are effective degrees
of freedom in a given context, and cannot be frozen for
ever. Otherwise, one would open a Pandora box containing too many new
exotic states. For instance, a light triquark $(\bar{s}ud)$, extrapolated 
naively from the pentaquark, would predict a bound
deuteron--$\overline{\Omega}{}^+$ with baryon number $B=1$ and
strangeness $S=3$. Similarly, a low mass $(cs)$ diquark \cite{Maiani:2004vq}
could suggest the possibility of a bound dibaryon with charm $C=3$ and
strangeness $S=-3$ below the threshold $(ccc)+(sss)$.
Already in 1972, Frederiksson and J\"andel  \cite{Fredriksson:1981mh} mentioned that the diquark model could lead to a ``demon-deuteron'' state $(ud)^3$. Though their paper has an erroneous statement about the quantum numbers of a three-boson system with an antisymmetric colour wave function, their warning remains.\footnote{A correspondance with H.J.\ Lipkin on this subjet is gratefully acknowledged.}

\subsection{Molecules}
It is regularly rediscovered that the Yukawa mechanism of nuclear
forces is by no means restricted to the two-nucleon systems and
presumably holds for any pair of hadrons containing light quarks.
Already Fermi and Yang noticed the possibility of a strong
attraction between a nucleon and an antinucleon, leading Shapiro, 
Dover and others to speculate about quasi-nuclear bound states as another approach
to baryonium \cite{Fermi:1959sa}.

More recently, several authors  noticed the presence of attractive forces
in some partial waves of the $D\overline{D}{}^*+\mathrm{c.c}$ and
$DD^*$ systems, mainly due to pion exchange. The potential is weaker
than for proton--neutron, but is experienced by heavier particles, and
 can give rise to binding. The former system,
$D\overline{D}{}^*+\mathrm{c.c}$, is perhaps seen in the $X(3872)$ \cite{Choi:2003ue},
which is just at the threshold \cite{Swanson:2003tb}.
 The latter, $DD^*$, would correspond
to an exotic meson  \cite{Manohar:1992nd} of charm $C=2$, and is also predicted in pure
quark-model calculations, due to the flavour independence of confining
forces, which favours $(QQ\bar{q}\bar{q})$ with respect to
$(Q\bar{q})+(Q\bar{q})$ if the mass ratio $m(Q)/m(q)$ is large enough
\cite{Ader:1981db}.

The case of baryons has been considered by Julia-Diaz and Riska \cite{Julia-Diaz:2004rf}
who found the possibility of bound states of charmed or multicharmed
baryons. For instance, two of the double charmed baryons \cite{Russ:2005uy} presented at
this Conference by J.~Russ might form a 
$[(ccq)-(ccq)]$ state stable against spontaneous dissociation, but perhaps above the lower threshold
$(ccc)+(cqq)$. This means that an entire new domain of nuclear physics
awaits discovery, since, ``nuclei'' made of several charmed baryons can be envisaged.

In these calculations, one often ends with a two-hadron interaction
which is marginally attractive enough to achieve two-body binding,
depending on uncontrollable details of the model. Here, the phenomenon
of Borromean binding comes to the rescue. Three-body bound states can 
exists whose two-body subsystems are unstable against dissociation \cite{Richard:2003nn}.

\subsection{Chiral dynamics}
Chiral dynamics offers a rigorous and self-consistent framework to study phenomenologically  light-quark physics at low energy. The successes are remarkable. 
 
An analysis of the underlying symmetry patterns led to interesting results for hadron spectroscopy, with  the possibility of low-lying scalar states, seen as partners of the ground-state pseudoscalars, and to a restoration of parity doublets for  highly-excited states. The most dramatic prediction is that of an antidecuplet of light baryons above the known octet (N, $\Lambda$, \dots) and decuplet ($\Delta$, \dots, $\Omega^-$). The $\theta^+(1540)$ and $\Xi^{-Ñ}(1870)$ are natural candidates for the exotic sector of this $\overline{10}$ multiplet \cite{Praszalowicz:2005tw}. Unfortunately, the $\theta^+(1540)$ is not seen in most of the high-energy experiments with excellent particle identification and acknowledged record in spectroscopy, as well as in recent measurements at Jlab, where earlier experiments gave positive results. 
 The $\Xi^{--}$ was claimed only by a fraction of a single collaboration, and never confirmed elsewhere and hence its evidence is even weaker. For a recent review of the situation on pentquarks, see, e.g., Kabana \cite{Kabana:2005tp}.

\subsection{Chromomagnetism}
The chromomagnetic interaction
\be
H_\mathrm{cm}=-\sum_{i<j} C_{ij}\tilde{\lambda}^c_i.\tilde{\lambda}^c_j\,\vec{\sigma}_i.\vec{\sigma_j}~.
\ee
though challenged by models based on instanton-induced interaction, or by spin--flavour dynamics, offers a convincing explanation of the hyperfine splittings of ordinary hadrons~\cite{Jaffe:1999ze}. Thus Hamiltonian, was very instrumental in demonstrating the possibility of low-lying $(q^2\bar{q}^2)$ states, to explain the presence of supernumerary scalar states of low mass, and in exhibiting the occurrence of coherent attractive forces in selected spin-flavour multiquark configurations.

The chromomagnetic model has been extensively studied, but it can still reveal some suprises  for exotic configurations, provided flavour-symmetry breaking is properly accounted for \cite{Hogaasen:2005}.

\section{Outlook}
A net revival of interest has been observed in the domain of
spectroscopy.  This takes now a larger fraction of the analysis effort
in multipurpose experiments with big detectors at major collider
facilities. Estimates of multiquark masses is not restricted to
potential models using somewhat ad-hoc colour dependence for the
interquark forces. Lattice QCD and QCD sum rules, for instance, are
now used to make predictions as a non-trivial extension of the previous
works on ordinary mesons and baryons.

It is hoped that  this
activity will survive the fashion for the pentaquark.  New sectors remain
to be scanned, in particular those mixing heavy quarks and light
quarks, beyond the most accessible hidden charm sector, and this involves intensive analysis.

\section*{Acknowledgements}
I would like to thank the organisers, M.~Hagenauer, B.~Nicolescu and
J.~Tran Thanh Van, for the stimulating atmosphere of this meeting,
several colleagues for fruitful discussions, in particular J.~Russ and
Tai T.~Wu, and M.~Asghar for comments on the manuscript.

\begin{small}

\end{small}

\begin{thebibliography}{99}
%
\bibitem{Martin:2003rr}
  A.~Martin and J.~M.~Richard,
  CERN Cour.\  {\bf 43N3} (2003) 17;
%
  T.~Barnes, T.~E.~Browder and S.~F.~Tuan,
  arXiv:hep-ph/0408081.
%
\bibitem{Giles:1976hh}
  R.~Giles and S.~H.~H.~Tye,
  Phys.\ Rev.\ Lett.\  {\bf 37} (1976) 1175;
%
D.~Horn and J.~Mandula,
Phys.\ Rev.\ D {\bf 17} (1978) 898;
%
P.~Hasenfratz, R.~R.~Horgan, J.~Kuti and J.~M.~Richard,
Phys.\ Lett.\ B {\bf 95} (1980) 299;
%
  T.~Barnes, F.~E.~Close, F.~de Viron and J.~Weyers,
   Nucl.\ Phys.\ B {\bf 224} (1983)  241;
%
  A.~Le Yaouanc, L.~Oliver, O.~Pene, J.~C.~Raynal and S.~Ono,
  Z.\ Phys.\ C {\bf 28} (1985) 309.
%

\bibitem{Abe:2004zs}
  K.~Abe {\it et al.}  [Belle Collaboration],
  Phys.\ Rev.\ Lett.\  {\bf 94} (2005) 182002.

\bibitem{Maiani:2004vq}
L.~Maiani, F.~Piccinini, A.~D. Polosa and V.~Riquer,
\newblock Phys. Rev. {\bf D71}, 014028 (2005); 
L.~Maiani, V.~Riquer, F.~Piccinini and A.~D. Polosa,
\newblock Phys. Rev. {\bf D72}, 031502 (2005). 

\bibitem{Close:2005iz}
  F.~E.~Close and P.~R.~Page,
  Phys.\ Lett.\ B {\bf 628} (2005) 215.

\bibitem{Kou:2005gt}
E.~Kou and O.~Pene,
\newblock hep-ph/0507119.

\bibitem{Jaffe:2003sg}
  R.~L.~Jaffe and F.~Wilczek,
  Phys.\ Rev.\ Lett.\  {\bf 91} (2003) 232003;
See, also,
  R.~L.~Jaffe,
  Phys.\ Rept.\  {\bf 409} (2005) 1.

\bibitem{Karliner:2003dt}
  M.~Karliner and H.~J.~Lipkin,
  Phys.\ Lett.\ B {\bf 575} (2003) 249.

\bibitem{Choi:2003ue}
  S.~K.~Choi {\it et al.}  [Belle Collaboration],
  Phys.\ Rev.\ Lett.\  {\bf 91}, 262001 (2003).

\bibitem{Aubert:2005rm}
  B.~Aubert {\it et al.}  [BABAR Collaboration],
  Phys.\ Rev.\ Lett.\  {\bf 95}, 142001 (2005).

\bibitem{Fredriksson:1981mh}
  S.~Fredriksson and M.~J\"andel,
  Phys.\ Rev.\ Lett.\  {\bf 48}, 14 (1982).

\bibitem{Fermi:1959sa}
  E.~Fermi and C.~N.~Yang,
  Phys.\ Rev.\  {\bf 76} (1949) 1739.
  I.~S.~Shapiro,
  Phys.\ Rept.\  {\bf 35} (1978) 129.
  C.~B.~Dover, T.~Gutsche and A.~Faessler,
  Phys.\ Rev.\ C {\bf 43} (1991) 379.

\bibitem{Swanson:2003tb}
E.~S. Swanson,
\newblock Phys. Lett. {\bf B588}, 189 (2004); 
N.~A. Tornqvist,
\newblock Phys. Lett. {\bf B590}, 209 (2004); 
M.~B. Voloshin,
\newblock Phys. Lett. {\bf B604}, 69 (2004); 
  E.~Braaten and M.~Kusunoki,
  Phys.\ Rev.\ D {\bf 72} (2005) 054022.

\bibitem{Manohar:1992nd}
  A.~V.~Manohar and M.~B.~Wise,
  Nucl.\ Phys.\ B {\bf 399} (1993) 17.

\bibitem{Ader:1981db}
  J.~P.~Ader, J.~M.~Richard and P.~Taxil,
  Phys.\ Rev.\ D {\bf 25} (1982) 2370;
  D.~Janc and M.~Rosina,
  Few Body Syst.\  {\bf 35}, 175 (2004);
  J.~M.~Richard and F.~Stancu,
  arXiv:hep-ph/0511043.

\bibitem{Julia-Diaz:2004rf}
  B.~Julia-Diaz and D.~O.~Riska,
  Nucl.\ Phys.\ A {\bf 755}, 431 (2005).

\bibitem{Russ:2005uy}
  J.~S.~Russ,
  Nucl.\ Phys.\ A {\bf 755}, 180 (2005).

\bibitem{Richard:2003nn}
  J.~M.~Richard, in the Book in honor of V.B.~Belyaev's 70th birthday, Dubna (2003),
  arXiv:nucl-th/0305076.



\bibitem{Praszalowicz:2005tw}
  M.~Praszalowicz and K.~Goeke,
  Acta Phys.\ Polon.\ B {\bf 36}, 2255 (2005).

\bibitem{Kabana:2005tp}
  S.~Kabana,
  J.\ Phys.\ G {\bf 31}, S1155 (2005).

\bibitem{Jaffe:1999ze}
  R.~L.~Jaffe,
  arXiv:hep-ph/0001123.

\bibitem{Hogaasen:2005}
H.~Hogaasen and P. Sorba, private communication and work in progress with F.~Buccella and J.-M. Richard.

\end{thebibliography}
\end{document}